\documentclass[article,usenatbib]{mn2e}  
\usepackage{graphicx,epsfig,psfig}  
\usepackage{amsmath}
\usepackage{amssymb}
\usepackage{kellymacros2} 
\usepackage[usenames]{color}

\long\def\Ignore#1{\relax}

\title[KIC\,8164262: Observations of Resonant Locking]{KIC
8164262: a Heartbeat Star Showing Tidally Induced Pulsations with Resonant Locking}

\author[Hambleton et~al.]{K. Hambleton$^{1}$\thanks{Email: kelly.hambleton@villanova.edu}, J. Fuller$^{2,3}$, S. Thompson$^{4}$, A. Pr\v{s}a$^{1}$,  D. W. Kurtz$^{5}$,\newauthor A. Shporer$^{6}$, H. Isaacson$^{7}$, A. W. Howard$^{8}$, M. Endl$^{9}$, W. Cochran$^{9}$, S. J. Murphy$^{10,11}$\\
$^{1}$Department of Astrophysics and Planetary Science, Villanova University, 800 East Lancaster Avenue, Villanova, PA 19085, USA\\ 
$^{2}$Kavli Institute for Theoretical Physics, Kohn Hall, University of California, Santa Barbara, CA 93106, USA\\
$^{3}$TAPIR, Walter Burke Institute for Theoretical Physics, Mailcode 350-17
California Institute of Technology, Pasadena, CA 91125\\
$^{4}$SETI Institute/NASA Ames Research center, Moffett Field, CA 94035, USA\\
$^{5}$Jeremiah Horrocks Institute, University of Central Lancashire, Preston, PR1~2HE\\
$^{6}$Division of Geological and Planetary Sciences, California Institute of Technology, Pasadena, CA 91125, USA\\
$^{7}$Astronomy Department, University of California, Berkeley, CA 94720, USA\\
$^{8}$California Institute of Technology, Pasadena, CA, 91125, USA\\
$^{9}$Department of Astronomy and McDonald Observatory, University
of Texas at Austin, 2515 Speedway, Stop C1400\\
$^{10}$Sydney Institute for Astronomy (SIfA), School of Physics, The University of Sydney, NSW 2006, Australia\\
$^{11}$Stellar Astrophysics Centre, Department of Physics and Astronomy, Aarhus University, DK-8000 Aarhus C, Denmark
}
 
\begin{document} 
\date{Accepted}  
\pagerange{\pageref{firstpage}--\pageref{lastpage}} \pubyear{2017}  
 
\maketitle  
\begin{abstract}
We present the analysis of KIC\,8164262, a \hb\ star with a high-amplitude ($\sim$1\,mmag), tidally resonant pulsation (a mode in resonance with the orbit) at 229 times the orbital frequency and a plethora of tidally induced g-mode pulsations (modes excited by the orbit). The analysis combines {\it Kepler} light curves with follow-up spectroscopic data from the Keck telescope, KPNO (Kitt Peak National Observatory) 4-m Mayal telescope and the 2.7-m telescope at the McDonald observatory. We apply the binary modelling software, \ph, to the \kep\ light curve and radial velocity data to determine a detailed binary star model that includes the prominent pulsation and Doppler boosting, alongside the usual attributes of a binary star model (including tidal distortion and reflection). The results show that the system contains a slightly evolved F star with an M secondary companion in a highly eccentric orbit ($e=0.886$). We use the results of the binary star model in a companion paper \citep{Fuller2017} where we show that the prominent pulsation can be explained by a tidally excited oscillation mode held near resonance by a resonance locking mechanism.
\end{abstract}


\section{Introduction}

Heartbeat stars are eccentric ellipsoidal variables (where the majority have eccentricities \,$\gtrsim$\,0.3) that undergo strong tidal interactions at the time of periastron passage, relative to the rest of the orbit. A consequence of these tidal interactions is that (for both components) the stellar cross-section changes shape, and the temperature across the stellar surface varies due to reflection and gravity darkening. These variations appear in the light curve primarily during periastron and the morphology of the variation depends on the eccentricity (and, in some cases, reflection). Heartbeat stars were initially identified as a separate class of binary star by \citet{Thompson2012} based on their unusual light curve morphology, specifically, their prominent periastron variations. Prior to this classification, there were two obvious cases in the literature: HD 174884 and KOI-54, which were reported by \citet{Maceroni2009} and \citet{Welsh2011}, respectively. Following this, many more were identified, the majority of which were found with the \kep\ satellite: KIC\,4544587 by \citet{Hambleton2013}; 17 red giant \hb\ stars by \citet{Beck2014}; KIC\,10080943 by \citet{Schmid2015}; six by \citet{Smullen2015} and 19 by \citet{Shporer2016}\footnote{\sf http://web.gps.caltech.edu/$\sim$shporer/heartbeatstars/}. The most up-to-date and extensive list of \kep\ \hb\ stars, containing \hbno\ objects, has been published by \citet{Kirk2015} and can be found at the \kep\ eclipsing binary website\footnote{\sf http://keplerEBs.villanova.edu}. Heartbeat stars have also been identified using other missions including seven with the Optical Gravitational Lensing Experiment, OGLE \citep{Nicholls2012}; one with CoRoT by \citet{Hareter2014}; and one discovered using {\sc most} and followed up with the CHARA array \citep{Richardson2016}. 

\begin{figure*} 
\hfill{} 
\centering
\includegraphics[width=\hsize]{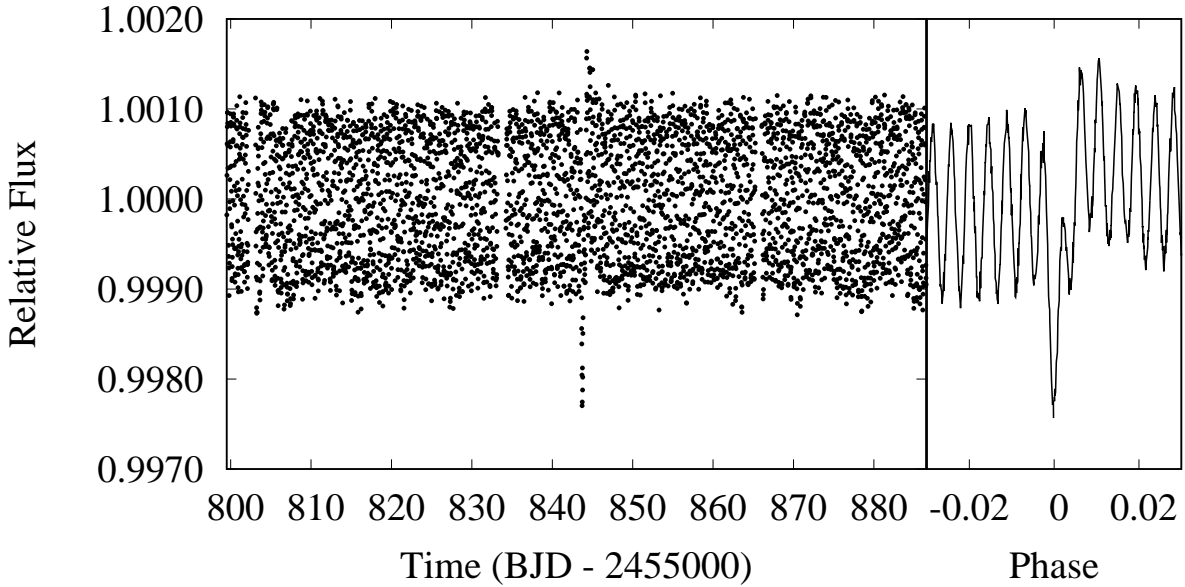}  
\small\caption{Left panel: The observed \kep\ light curve of KIC\,8164262 for a single orbit of 87.45\,d during Quarter 9. Right panel: a magnified region of the phase-binned \kep\ light curve from Quarters 0--17, containing the ellipsoidal variation at phase zero and showing the prominent pulsation variations. } 
\label{fig:lc} 
\hfill{} 
\centering
\end{figure*} 

Heartbeat stars are a diverse collection of objects, some of which display additional interesting characteristics such as solar-like oscillations \citep{Beck2014}, rapid apsidal motion \citep{Hambleton2013,Hambleton2016b} and tidally induced pulsations \citep{Welsh2011}. Tidally induced pulsations, initially theorised by \citet{Zahn1975}, \citet{Goldreich1989} and \citet{Witte2002}, are pulsations driven by the varying tidal forces that occur as the stars orbit each other. They were hypothesised to cause the circularisation of binary star orbits and the spin-up of the stellar components, although, until \kep\ their presence had only been identified in HD\,174884 \citep{Maceroni2009}. Thanks to \kep, we now have a plethora of objects with tidally induced pulsations; approximately 20\% of the current \kep\ \hb\ star sample show tidally induced pulsations, providing us with a range of pulsation frequencies ($\lesssim$\,10\,\cd) and amplitudes ($\lesssim$\,1\,mmag) to investigate. Interestingly, the phase of each tidally induced pulsation, relative to the phase of periastron, is determied by the azimuthal order of the mode \citep{Burkart2012}. To determine the azimuthal order, however, extremely high precision is required for both the argument of periastron and the pulsation phases.

Tidally induced modes are forced to oscillate at integer multiples of the orbital frequency, and their amplitude is determined by the difference between the tidal forcing frequency and a mode's natural frequency (Kumar et al. 1995). Very large amplitude tidally induced modes are unlikely because they require a near exact resonance between tidal forcing and one of the star's natural mode frequencies. Interestingly, however, with \kep\ we have observed thirteen objects that exhibit a dominant high-amplitude pulsation mode that appears to be in resonance with the binary star orbit (these are classified as resonant modes). KIC\,8164262, the focus of this work, is an extreme case with a dominant, high-amplitude ($\sim$1\,mmag) mode. Here we propose the theory of resonance locking, hypothesised by \citet{Witte1999}, \citet{Witte2001}, \citet{Fuller2012}, \citet{Burkart2012} and \citet{Burkart2014}, as the mechanism for KIC\,8164262 maintaining its resonant mode \citep{Zahn1975,Zahn1977}.

The proposed mechanism of resonance locking achieves this by locking the evolution of the binary star orbital period with the evolution of the eigenmodes as the stars evolve. Resonance locking can occur because a star's oscillation mode frequencies change due to stellar evolution. The tidal forcing frequencies also change due to orbital evolution caused by tidal dissipation, at a rate that depends on how close mode frequiences are to resonance. Near resonances, these rates of evolution can be equal to one another, allowing for a stable equilibrium in which the star and orbit evolve in tandem such that a tidally forced mode remains near resonance \citep{Witte1999}. Consequently, rather than passing through resonance, tidally induced pulsations can be locked in resonance and are more likely to be observed at large amplitudes.


A more extensive theoretical discussion of resonance locking, both generally and for the case of KIC\,8164262, is described in the companion paper by Fuller et al. (2017, submitted; hereafter F17), who provide theoretical models showing that the prominent pulsation in KIC\,8164262 aligns with the predictions of the resonant locking mechanism, given the fundamental stellar parameters (mass and radius of the primary component) we have determined. 

Here we present the observations and characterization of KIC\,8164262. In \S\ref{sec:obs} we describe the ground- and space-based observations; in \S\ref{sec:binary} we outline the detailed binary model of KIC\,8164262; in \S\ref{sec:pulse} we discuss the pulsations; and in \S\ref{sec:summary} we discuss and summarise our findings. 

\section{Observations}
\label{sec:obs}


KIC\,8164262 (see Fig.\,\ref{fig:lc}) was initially identified as a heartbeat star by the Planet Hunters \footnote{\sf https://www.planethunters.org} and forwarded to the Kepler Eclipsing Binary Working Group where it was subsequently added to the Kepler Eclipsing Binary Catalog \citep{Kirk2015}. It was also previously identified in the \citet{Slawson2011} catalog, but with an erroneous period based on the prominent pulsation rather than the binary features. KIC\,8164262 was selected for detailed study primarily due to its high amplitude resonantly excited mode, which makes it a strong candidate for resonant locking.

\subsection{\kep\ Data}

The \kep\ telescope \citep{Borucki2010, Gilliland2010,Batalha2010} observed KIC\,8164262, which has a \kep\ magnitude of Kp\,=\,13.36, nearly continuously for 1470.5\,d or 17 Quarters. The observations of KIC\,8164262 were obtained using the  long cadence (hereafter LC) data mode at a mean sampling rate of 29.4244\,min. All observations were obtained from the Mikulski Archive for Space Telescopes and were a part of Data Releases 21--23 \citep{DRN21, DRN22, DRN23,DRN25}.

To create a time series of the relative flux variations of KIC\,8164262, we used barycentric times as reported in the {\sc time} column and the fluxes reported in the {\sc pdcsap\_flux} data column of the \kep\ data files. These data have been processed through the \kep\ pipeline \citep{DataProcessingHandbook}, including the PDC (Presearch Data Conditioning) module of the pipeline, which uses a Bayesian, multi-scale principal component analysis to remove common features from the time series \citep{Smith2012,Stumpe2012,Stumpe2014}. We then fitted a low order ($<$4) polynomial to the time series of each Quarter individually. Our final light curve was created by dividing by this fit to yield the fractional variation in the flux. 
 
As each \kep\ pixel is 4\,$\times$\,4\,arcsec, it is possible that some contamination may occur within the photometric mask in the form of light from an additional object. The maximum reported statistical contamination value for KIC\,8164262 is 0.002 for all observed Quarters on a scale of 0 to 1, where 0 implies no contamination and 1 implies complete contamination of the CCD pixels by other stars in the aperture. The low value of 0.002 suggests that KIC\,8164262 suffers minimally from third light, if at all. To assess the flux incident on each individual pixel we used pyKE \citep{Still2012} to generate the per-pixel light curve plots and examine the flux distribution over the newly defined masks. From this we visually confirmed that no other source is contaminating our observations.  

\subsection{Period Determination}
\label{sec:period}

Period analysis was performed to identify the orbital period of the binary 
system. The analysis was done on all data, Quarters, using {\sc kephem} \citep{Prsa2011}, an interactive package with a graphical user interface that incorporates 3 period finding methods: Lomb-Scargle (LS; Lomb 1976; Scargle 1982),\nocite{Lomb1976, Scargle1982} Analysis of Variance (AoV; Schwarzenberg-Czerny 1989),\nocite{Schwarzenberg-Czerny1989} and Box-fitting Least Squares (BLS; Kov{\'a}cs et al. 2002),\nocite{Kovacs2002} as implemented in the {\sc vartools} package (Hartman et al. 1998)\nocite{Hartmann1998}. Using {\sc kephem}, the period and time of the deepest minimum of the ellipsoidal variation were found interactively. The ephemeris was found to be:\newline 
\newline 
Min\Rmnum{1} = BJD 2455668.82920+87.4549(6) $\times$ E\newline
\newline 
\noindent 
where Min\Rmnum{1} refers to the deepest minimum of the ellipsoidal variation (identified by eye) counted from the centre of the data set. The values in the parenthesis gives the 1$\sigma$ uncertainty in the last digit. The period uncertainty was obtained by applying an adaptation of the Period Error Calculator algorithm of \citet{Mighell2013}, as specified by \citet{Kirk2015}. 

\subsection{Ground Based Spectroscopy}

We obtained three sets of spectra: fifteen spectra from the HIRES spectrograph on the Keck telescope, Mauna Kea; two spectra using the Tull spectrograph on the 2.7-m telescope at McDonald Observatory; and two spectra using the Echelle Spectrograph on the 4-m Mayall telescope, Kitt Peak National Observatory (KPNO). The object was determined to be an single lined spectroscopic binary system, as described in \S\ref{sec:todcor}. The radial velocities derived from the three sets of observations are reported in Table\,\ref{tab:rvs}. 

Keck observations were taken with the standard setup of the California Planet Search \citep{Howard2010} over the course of three months, beginning in 2015 May. Exposure times were between 120 and 180\,s and each spectrum has a signal-to-noise (SNR) of ~52 per resolution element at 5500\,\AA\ with a resolution of ~60\,000. In order to calculate the systemic radial velocity, we utilize the telluric A and B absorption features that fall on 7594--7621\,\AA\ and 6867--6884\,\AA, respectively. Using the method from \citet{Chubak2012}, the positions of the primary star's spectral lines were measured relative to the telluric features. The positions of the spectral lines were converted into radial velocities and an offset was applied to place the relative radial velocities on the standard scale used by \citet{Nidever2002} and \citet{Latham2002}.
 
We further observed KIC 8164262 with the Tull Coud\'e Spectrograph mounted on the Harlan J. Smith 2.7-m Telescope \citep{Tull1995} at McDonald Observatory. The Tull spectrograph covers the entire optical spectrum at a resolving power of R\,=\,60\,000. We collected 2 spectra in 2015 August using exposure times of 800~s. The resulting spectra have SNR ratios from 23 to 26 per resolution element at 5650\,\AA. For each target visit we also obtained a spectrum of HD~182488, the radial velocity standard star used for the \kep\ field, which we used to measure absolute radial velocities by cross-correlating the target star's spectra with this standard star spectrum. 

The KPNO observations were taken in sets of back-to-back exposures on 2013 May 29--30 (900\,s each) and 2013 June 17--18 (750\,s each). Calibration exposures were taken using a ThAr lamp prior to each exposure. Using the echelle spectrograph, a wavelength coverage of $4600-9050$\,\AA\ was obtained with a resolving power of R\,$\sim$20\,000. The signal-to-noise ratio obtained was $\sim$40 per resolution element. The data were reduced using the {\sc iraf} (Image Reduction and Analysis Facility) software package \citep{Tody1986, Tody1993}.

\begin{figure} 
\hfill{} \hspace{-0.9cm}
\includegraphics[width=10cm]{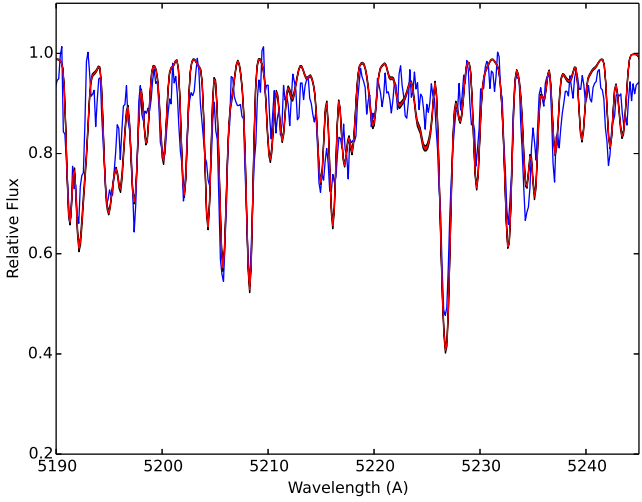}  
\small\caption{A section of a KPNO spectrum from 5190--5245\,\AA\ (blue line) with the best fit model (red line) using \tdcr\ combined with \mcmc. The entire fitted region extends from 4800\,\AA\ to 6750\,\AA.} 
\label{fig:spectra} 
\hfill{} 
\end{figure}

\begin{figure} 
\hfill{} 
\centerline{\includegraphics[width=9cm]{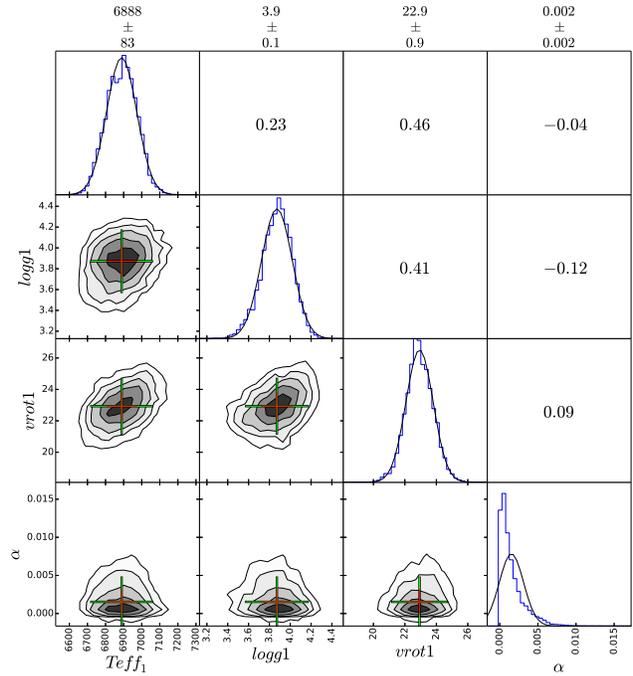}}  
\small\caption{Posterior probability distribution functions produced by applying \tdcr\ combined with \mcmc\ to the spectra obtained using the Mayall telescope at KPNO. Lower left subplots: two dimensional cross-sections of the posterior probability distribution functions. The crosses show the 1$\sigma$ (red) and 2$\sigma$ (green) uncertainties, and are centred on the minima. Diagonal subplots from top left to bottom right: histograms displaying the probability distributions of the effective temperature, $T_{\rm eff}$ (K); the surface gravity, $\log g$ (dex); and the projected rotational velocity, $v \sin i$ or $vrot$ ($\kms$); for the primary component, and $\alpha$, the fractional light contribution, $f_2/(f_1+f_2)$, where $f_1$ and $f_2$ are the light contributions of the primary and secondary, respectively. Upper right subplots: the correlations for the two-dimensional cross-sections mirrored in the diagonal line where 1 is a direct correlation and -1 is a direct anti-correlation. The values above the plot give the mean value and one sigma uncertainty for each parameter, based on the fitted Gaussians. For the $\alpha$ parameter, the peak of the distribution is at $\sim$0 and thus the normal distribution is not a good approximation. For this reason, the fitted value above the plot is not representative.}
\label{fig:tdcr} 
\hfill{} 
\end{figure} 
 
\subsubsection{Deriving Fundamental Parameters and Radial Velocities from the KPNO Spectra}
\label{sec:todcor}

The radial velocity data from the KPNO observations were generated using the 2-D cross-correlation technique as implemented in \tdcr\ \citep{Zucker1994} combined with the {\sc python} implementation of {\sc emcee}, an affine invariant version of Markov chain Monte Carlo ({\sc mcmc}) method, proposed by \citet{Goodman2010} and implemented by \citet{DFM2013}. By combining these software packages, we were able to simultaneously obtain the posteriors of the fundamental parameters: effective temperature, $T_{\rm eff}$, projected rotational velocity, $v$\,$\sin$\,$i$, and gravity,  $\log g$; and obtain radial velocity distributions (distributions of possible radial velocities based on the range of possible spectral models and \tdcr\ uncertainties; \citet{Hambleton2016a}). Although \tdcr\ is designed for double lined spectroscopic binaries, by applying it to KIC\,8164262, we were able to marginalize over the parameters relating to the unconstrained secondary component and thus propagate them forward.

\mcmc\ is used to explore the binary star parameter space using a set of Markov chains, in this case 128. These chains begin with random distributions based only on their prior probability distribution functions. For both components, we provided uniform priors for $T_{\rm eff}$, $v \sin i$ and $\log g$: 5000--8500\,K, 0--100\,km\,s$^{-1}$ and 2--5\,dex, for the primary component; and 3000--5000\,K, 0--100\,km\,s$^{-1}$ and 4--5\,dex for the invisible secondary component. We also fitted the fractional light contribution, $\alpha = f_2/(f_1+f_2)$, where $f_1$ and $f_2$ are the light contributions of the primary and secondary, respectively and provided a prior of 0--0.1. While the results of the secondary component are inconclusive due to the low light contribution ($<$\,0.5 per cent), we marginalized (integrated) over all possible values of the secondary star's atmospheric parameters to avoid biasing our results by selecting a specific spectrum. Consequently, the uncertainty stemming from the unconstrained secondary star parameters is propagated forward.

At each step, two spectra were generated (one for each component), from a grid of templates that are synthesized with {\sc spectrum} \citep{Gray1999} using \citet{Castelli2004} model atmospheres. The radial velocities of the primary component were then determined by applying \tdcr\ to the observations using the templates (adjusted to  account for their light contributions). The $\chi^2$ value was determined between the shifted, synthetic spectra and the observed spectra. We specified a global per-point uncertainty for the two spectra of $\sigma$\,=\,0.03, which we determined by considering the noise level of the spectra. Each $\chi^2$ value was then multiplied by -0.5 to obtain the log likelihood for each observed spectrum and the results were summed over all spectra.

At each iteration the radial velocities and associated errors produced by \tdcr\ were also stored. The radial velocity distributions were then determined by combining the \tdcr\ radial velocity values and errors with the spread caused by the uncertainty in the model spectra. The outcome is a distribution of radial velocities that is marginalized over the model spectra and includes the uncertainties from \tdcr. During the initial burn-in time, the Markov chains converge towards their maximum likelihood value. The statistics of a large number of iterations ($\sim$10\,000, excluding the burn-in time), provide posterior probability distributions for the model parameters. We applied this scheme to the two high resolution KPNO spectra of KIC\,8164262 using the spectral range of 4800--6750\,\AA. 

Due to its slow rotation (relative to stars above the \citet{Kraft1967} break), we anticipated that KIC\,8164262 would be a metal rich star. Consequently, we repeated the aforementioned spectral fitting for a range of metallicities: [Fe/H] = -0.2 to +0.5 in steps of 0.1. By comparing the log likelihoods in each case, we found that the metallicity of [Fe/H]\,=\,0.50\,$\pm$\,0.05 provided the best fit (where the \citet{Castelli2004} model libraries have a maximum metallicity of [Fe/H]\,=\,0.5). As we fit the entire spectrum, rather than specific lines, we cannot precisely infer the metallicity using this method, although we conclude that the spectrum of KIC\,8164262 is metal rich. Using this method we determined the KPNO radial velocities provided in Table\,\ref{tab:rvs} and found that KIC\,8164262 is a single-lined spectroscopic binary with the fundamental parameters listed in Table\,\ref{tab:fundamental}. The best-fit model to the spectrum is depicted in Fig.\,\ref{fig:spectra}. The posterior distributions of the spectral parameters are depicted in Fig.\,\ref{fig:tdcr} and are all Gaussian with the exception of the light ratio, which is consistent with $\sim$0, suggesting that the light from the secondary component is negligible. This demonstrates that our model is well constrained.

\begin{table}
\setlength{\tabcolsep}{2pt}
\centering
\caption[]{Original, unshifted radial velocities and their uncertainties for the primary component of KIC\,8164262. The spectral observations were taken using the echelle spectrograph on the 4-m Mayall telescope, Kitt Peak (KPNO), the HIRES  spectrograph on Keck and Cross-Dispersed Echelle Spectrograph on the 2.7-m telescope at the McDonald Observatory, Fort Davis.}
\begin{tabular}{l l r c l}
\hline
\multicolumn {2} {l}{Time (BJD)}&\multicolumn {3} {c} {RV1 (km\,s$^{-1}$)}\\\hline
Keck & & & \\\hline
2457151.059455     & &    $24.9$& $\pm$& 5.7 \\
2457202.892222     & &    $15.3$& $\pm$& 3.1 \\
2457228.978804     & &    $17.6$& $\pm$& 1.8 \\
2457237.073545     & &    $28.4$& $\pm$& 6.0 \\
2457239.988444     & &    $33.5$& $\pm$& 0.7 \\
2457241.068023     & &    $37.2$& $\pm$& 0.8 \\
2457243.031959     & &    $15.2$& $\pm$& 9.5 \\
2457244.791056     & &    $-9.8$& $\pm$& 0.5 \\
2457247.027160     & &    $-0.5$& $\pm$& 2.5 \\
2457255.883316     & &    $3.0$&  $\pm$& 0.9 \\
2417568.929067     & &    $23 $&  $\pm$& 4 \\
2417592.053308     & &    $36 $&  $\pm$& 7 \\
2417596.036900     & &    $-8 $&  $\pm$& 4 \\
2417599.999567     & &    $4.0$&  $\pm$& 0.9 \\
2417621.935110     & &    $16 $&  $\pm$& 2 \\
\hline                         
\multicolumn {4} {l}{KPNO}\\\hline
2456442.7814 & & 16.4& $\pm$&  2.1 \\  
2456461.7050 & & -2.3& $\pm$&  2.0 \\
\hline
\multicolumn {4}{l}{McDonald}\\\hline
2457242.6297& & 29.3 &$\pm$& 1.3 \\
2457250.8001& & 1.81 &$\pm$& 0.94 \\
\hline
\end{tabular}
\label{tab:rvs}
\end{table}

\begin{table}
\setlength{\tabcolsep}{2pt}
\centering
\caption[]{Fundamental parameters determined using \tdcr\ combined with {\sc emcee}. The method was applied to the spectral range 4800--6750\,\AA. }
\begin{tabular}{l l r c l}
\hline
\multicolumn {2} {l}{Parameters} & \multicolumn {3}{c}{Values}\\\hline 
$T_{\rm eff}$ (K)         & & 6890 &$\pm$& 80 \\
$\log\,g$ (dex)           & & 3.9  &$\pm$& 0.1\\
$v \sin i$ (km\,s$^{-1}$) & & 23   &$\pm$& 1\\
\hline
\end{tabular}
\label{tab:fundamental}
\end{table}

\section{Binary Star Model}
\label{sec:binary}

\subsection{Creating a Binary Model}
\label{sec:ph}

We used the binary modelling code {\sc phoebe} \citep{Prsa2005}, which is an extension of the Wilson-Devinney code \citep{Wilson1971,Wilson1979,Wilson2004}, to the light curve of KIC\,8164262. {\sc phoebe} combines the complete treatment of the Roche potential with the detailed treatment of surface and horizon effects such as limb darkening, reflection, and gravity brightening to derive an accurate model of the binary star. The implementation used here (version 1.0) relies on the Wilson-Devinney method of summing over the discrete rectangular surface elements that cover the distorted stellar surfaces. An accurate representation of the total observed flux and consequently a complete set of stellar and orbital parameters is then obtained by integrating over the visible elements. {\sc phoebe} incorporates all the functionality of the Wilson-Devinney code, but also provides an intuitive graphical user interface alongside many other extensions, including updated filters and bindings that enable interfacing between {\sc phoebe} and {\sc python}.

As modelling a large number of data points is computationally expensive, we elected to phase-bin the data for the purpose of binary modelling. This is appropriate for KIC\,8164262, as the binary features and tidally induced pulsation both repeat precisely every orbital cycle. Furthermore, KIC\,8164262 has no significant temporal variations that would affect the binned light curve, \ie\ apsidal motion, which would cause a change in the shape of the ellipsoidal variation as a function of time. We also note that this method significantly weakens the rotational signal due to spots, which is advantageous as we do not fit the rotation signal as part of our model (rather, we assumed pseudo-synchronous rotation based on this signal).

As the information content of the light curve peaks at the time of periastron passage, we did not bin the data between the phases -0.01 and 0.01 where we kept all the data points. At all other phases we binned the data into bins of 100 points, thus significantly reducing the number of data points in these regions. Rather than having discrete segments in the light curve with different cadences, we used a sigmoid function to bridge the number of data points between regions. By using this method, we avoided discrete jumps in the number of data points. Finally, we removed any obvious outliers ($\sim$100 data points) from the data by eye.

\begin{table} 
\caption{ 
\label{tab:fix} 
\small Fixed parameters and coefficients for the {\sc phoebe} model to 
the light and radial velocity curves for all available quarters. As the secondary component contributes an insignificant amount of light, the secondary parameters are also insignificant; however, the parameter values that we selected (based on estimates) are presented here for completeness.} 
\begin{center} 
\begin{tabular}{||l|r||} 
\hline 
Parameter & Value\\ 
\hline 
Orbital Period (d)                          & 87.4549\\ 
Time of primary minimum (BJD)               & 2455668.829\\ 
Primary $T_{\mathrm{eff}}$ (K), $T_{1}$     & 6890\\ 
Primary synchronicity parameter, $F$        & 29.2547\\
Primary bolometric albedo                   & 0.6\\  
Primary gravity brightening                 & 0.32\\  
Secondary $T_{\mathrm{eff}}$ (K), $T_2$     & 3500\\
Secondary radius (\Rsun), $R_2$             & 0.3\\ 
Secondary synchronicity parameter, $F$      & 29.2547\\
Secondary bolometric albedo                 & 0.6\\ 
Secondary gravity brightening               & 0.32\\ 
Third light                                 & 0.0\\ 
\hline 
\end{tabular} 
\end{center} 
\end{table} 

\begin{table} 
\hfill{} 
\caption{ 
\label{tab:free} 
\small Adjusted parameters and coefficients of the best-fit model to the light and radial velocity curves for the phased light curve data and all radial velocity measurements. The limb darkening coefficients were calculated using \ph. The passband luminosities were derived using the assumed temperature and radius of the secondary. The RV shift is a vertical shift applied to {\sc kpno} and McDonald radial velocities to shift them onto the standard scale used by \citet{Nidever2002} and $F$ is the stellar to orbital rotation rate. The fit was performed using MCMC methods and the values in the brackets denote the 1\,$\sigma$ uncertainties.}  
\begin{center} 
\begin{tabular}{||l|r||} 
\hline 
Parameter   & {Value}\\ 
\hline 
Mass ratio, $q$                             & 0.20(4)\\ 
Primary mass ($\Msun$), $M_1$               & 1.70(9)\\ 
Secondary mass ($\Msun$), $M_2$             & 0.36(2)\\ 
Primary radius (\Rsun), $R_1$               & 2.4(1)\\ 
Phase shift, $\phi$                         & 0.014(1)\\ 
Orbital eccentricity, $e$                   & 0.886(3)\\ 
Argument of periastron (rad), $\omega$      & 1.48(1)\\ 
Orbital inclination (degrees), $incl$       & 65(1)\\ 
Primary passband luminosity (\%), $L_1$              & 98.9(2)\\ 
Secondary passband luminosity (\%), $L_2$            & 1.1(1)\\ 
Semi-major axis (\Rsun), $sma$              & 106(2)\\
Primary $\log$\,$g$ (cgs),  $\log g1$       & 3.90(3)\\ 
Primary linear limb darkening coeff.        & 0.647\\
Secondary linear limb darkening coeff.      & 0.714\\
Primary logarithmic limb darkening coeff.   & 0.220\\
Secondary logarithmic limb darkening coeff. & 0.148\\
KPNO RV shift ($\kms$),                     & 2.7(1)\\
McDonald RV shift ($\kms$),                 & 0.01(2)\\
\hline
\end{tabular} 
\hfill{} 
\end{center} 
\end{table} 

The final binary star model was converged using a combination of \ph\ and \emcee. However, to understand our model parameters, we initially created a binary star model using \ph's GUI (graphical user interface). For this initial stage we prewhitened the primary pulsation from the light curve. When using the \ph\ GUI, we identified the parameters that significantly impact the light curve shape of KIC\,8164262. As this object is a single-lined non-eclipsing spectroscopic binary, this excludes the majority of parameters that pertain solely to the secondary component, with the exception of the upper limit on the secondary star's relative luminosity. To ensure that this was the case for the secondary radius, we computed the results with a radius of $R_2$\,=\,0.3\,\Rsun\ and $R_2 = 0.5$\,\Rsun\ and found that the results were identical within statistical uncertainties.

The parameters that were found to affect the binary star light and radial velocity curves are the eccentricity, inclination, argument of periastron, primary radius, primary gravity brightening exponent, luminosity ratio, mass ratio, semi-major axis and systemic velocity, where the phase shift is a convenience parameter that shifts the model horizontally to keep the minimum of the ellipsoidal variation at phase 0.0. As the gravity darkening exponent is degenerate with the primary star's radius, we elected to fix the gravity darkening exponent to 0.32, which is the theoretical value for stars with convective outer envelopes \citep{Lucy1967}, even when the envelope is very thin. As the secondary is small and cool, the amount of reflection in the light curve is negligible. As such, we elected to fix the albedo to the theoretical value for stars with convective envelopes, 0.6 \citep{Lucy1967}, for both components. A list of all the fixed parameters in our binary model can be found in Table\,\ref{tab:fix}.

\subsection{Parameter Space Sampling}

To create the final model we combined \ph\ with \mcmc\ to integrate the power of \ph\ as a binary modelling code with Bayesian inference. This was possible due to the recent update of \ph\ to include \python\ interfacing. We again elected to use \emcee, which is discussed in detail in \S\ref{sec:todcor}. In addition to the standard functionality of \ph, our models include the ability to fit Doppler boosting, as described by \citet{Bloemen2011}, and tidally induced pulsations. 

For KIC\,8164262 we elected to fit the high-amplitude prominent pulsation simultaneously with the binary star features. The signature of a tidally induced pulsation is a frequency that is a precise multiple of the orbital frequency. The prominent pulsation in KIC\,8164262 is 228.999(2)\,$\nu_{orb}$ which is equal to 229\,$\nu_{orb}$, given the uncertainty on the orbital period and pulsation frequency. In our model we fixed the frequency of the pulsation to the multiple of the orbital frequency and fitted the phase and amplitude to create a comprehensive binary star model. 

Doppler boosting is proportional to the radial velocity of the two stars and is the combined effect of shifting the stars' spectral energy distributions (relative to the \textit{Kepler}\ passband), aberration and an altered photon arrival rate. The net result of Doppler boosting is an increase in the observed flux from a star when it moves towards the observer, and a decrease when it moves away from the observer. It was predicted to be seen in the \textit{Kepler}\ data by \citet{Loeb2003} and \citet{Zucker2007}, and has been observed in several systems from ground-based data, as well as \textit{Kepler}\ and CoRoT light curves (see e.g. \citealt{Mazeh2010,van-Kerkwijk2010,Shporer2010,Bloemen2011,Shporer2016}. To determine the Doppler boosting coefficients, we used look-up tables, based on each component's effective temperature and $\log g$. These look-up tables take into account the spectrum of the star and the wavelength of the observations, and were computed from Kurucz 2004 model spectra \citep{Castelli2004}, using Eq.\,(3) of \citet{Bloemen2011}. The Doppler boosting contribution was estimated to be $B\sim$400\,ppm, which is significant given the peak-to-peak amplitude of the light curve is $\sim 4000$\,ppm. The calculation for Doppler boosting was performed at each iteration. 

\begin{figure}
\centerline{\includegraphics[height=7cm]{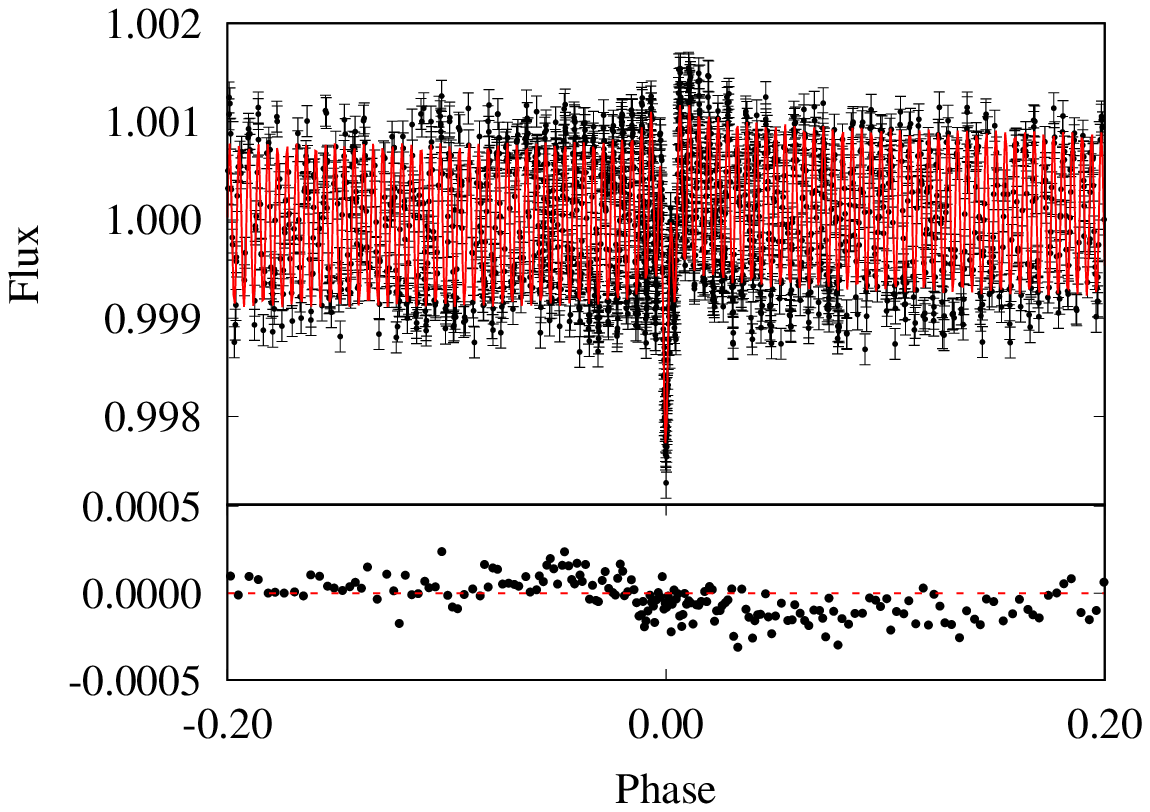}}  
\centerline{\includegraphics[height=7cm]{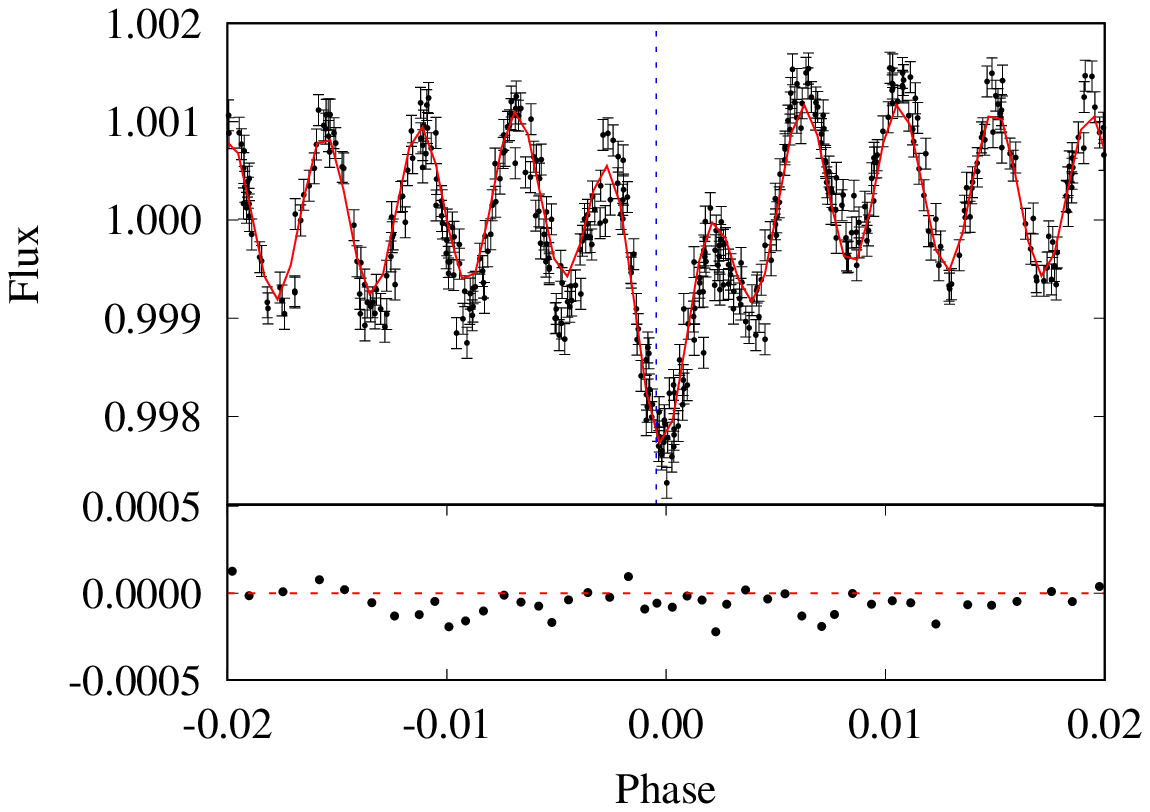}}  
\small\caption{Upper panel: The best-fit light curve model (red line) to the \kep\ data of KIC\,8164262 (black). The 1\,$\sigma$ uncertainties are denoted on the plot. The residuals of the best-fit model are provided below the model and have been binned for visual purposes. The red dashed line denotes zero flux. Lower panel: A magnified section of the periastron variation displaying the best-fit model. The blue dotted vertical line denotes the time of periastron passage, which is slightly offset from the observed time of minimum} 
\label{fig:lc_model} 
\end{figure}

In our model we restricted the log\,$g$ of the primary component to that determined from spectral fitting, $\log g1$\,=\,3.9\,$\pm$\,0.3 (where $\sigma = 0.1$). Consequently, at each iteration we calculated the primary star's gravitational potential (an input for \ph\ that is a proxy for the stellar radius), which is a function of the mass ratio, instantaneous separation, spin-to-orbital period ratio and log\,g. We calculated stellar luminosity, thus reducing the number of fitted parameters to twelve. Of these fitted parameters, eight are binary star parameters: the inclination, eccentricity, argument of periastron, phase shift, mass ratio, semi-major axis, systemic velocity and log\,$g$ of the primary component. Two pulsation parameters are the amplitude and phase of the high-amplitude, tidally-induced pulsation, and two are vertical radial velocity shifts to account for having radial velocity data from three different telescopes (these shifts were added to the original radial velocity values that are presented in Table\,\ref{tab:rvs}). These parameters were selected based on their significant contribution to the light curve. Other important parameters that were not fitted include: the primary effective temperature, which we fixed based on the spectral information; the period and zero point in time, which were fixed based on our period determination; the stellar rotation rate, which we fixed based on the stellar rotation signature in the light curve due to spots; and the aforementioned primary gravity darkening exponent, which we fixed to the theoretically determined value of 0.32.

\begin{figure*}
\hfill{} 
\centerline{\includegraphics[height=12cm]{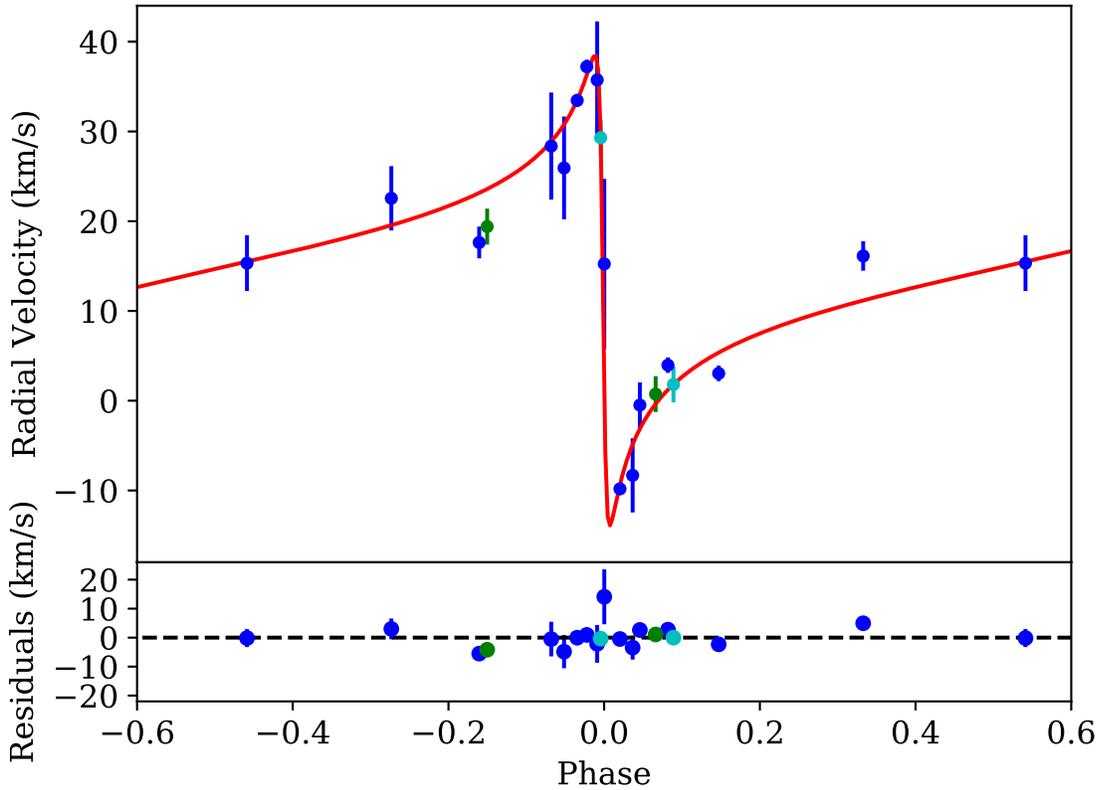}}  
\small\caption{Upper panel: The best-fit radial velocity curve (red) to the KPNO radial velocity points (green), Keck radial velocity points (blue) and McDonald radial velocity points (cyan). Vertical shifts, which were fitted simultaniously with the binary star model, have been applied to the McDonald and KPNO data to align the radial velocity points onto a single scale. The vertical shifts are attributed to the use of different telescopes. Bottom panel: the residuals of the best-fit model (note the change of scale). The error bars denote the one sigma uncertainty on the radial velocities.} 
\label{fig:rv_model}  
\hfill{} 
\end{figure*}

For each parameter we used a flat, uniform prior where the ranges were selected to be as large as possible without creating unphysical models, with the exception of log\,$g$, which we constrained to be within three sigma of the value obtained through spectral fitting. The likelihood was generated by multiplying the $\chi^2$ value from the light curve data by -0.5 and summing this with the $\chi^2$ value from the radial velocity data, again, multiplied by -0.5. Fig.\,\ref{fig:lc_model} depicts the model fit to the light curve. The light curve fit obtained is well constrained, as shown by the residuals presented in the lower panels. It can be seen that the amplitude of the pulsations surrounding the periastron variation are slightly underestimated. We believe that the comprehensive treatment of pulsations within the binary model framework would elevate this discrepancy. However, this is beyond the scope of this work. Fig.\,\ref{fig:rv_model} depicts the radial velocity curve (red line) and data.

\begin{figure*} 
\hfill{} 
\centerline{\includegraphics[height=\hsize]{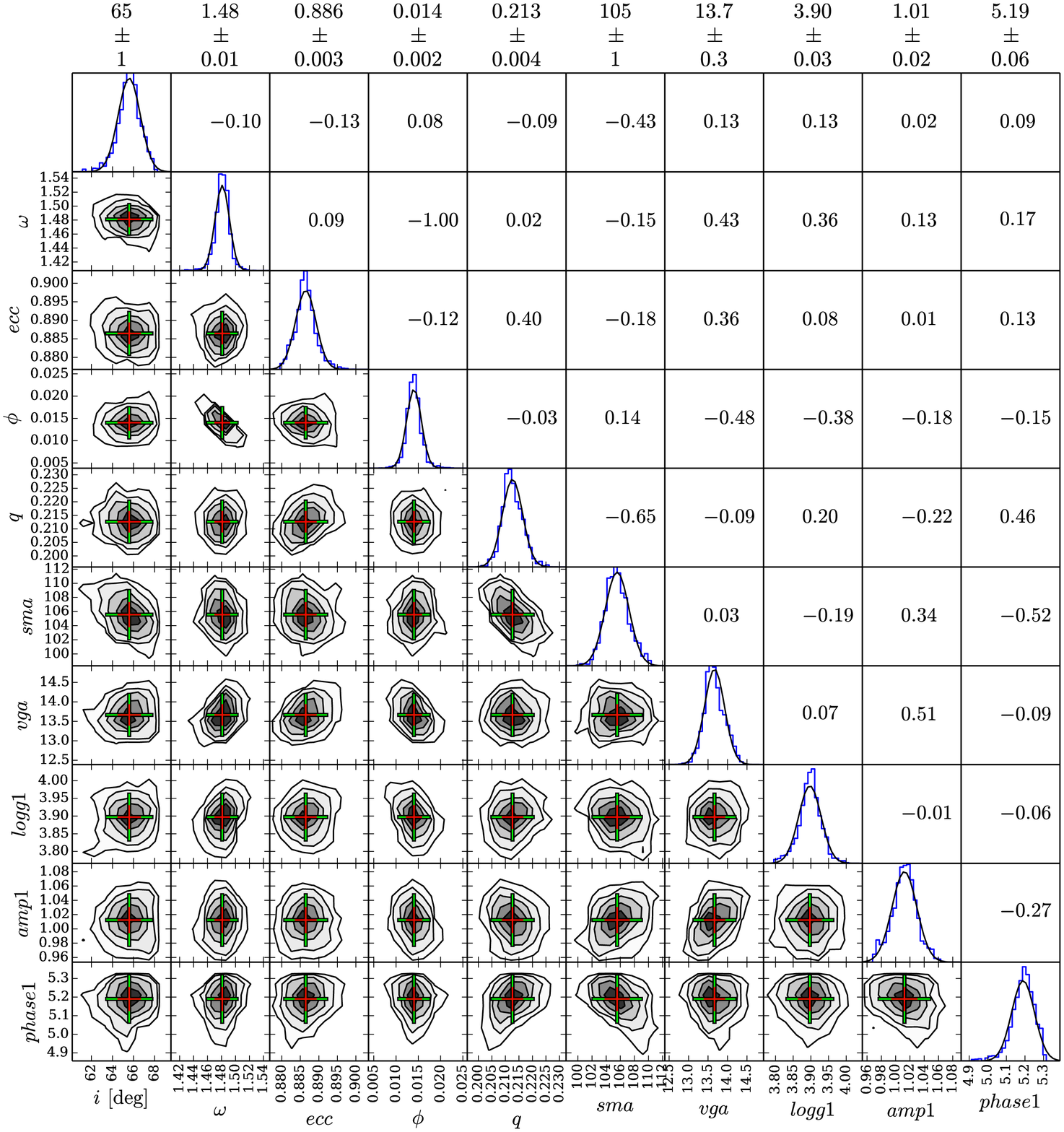}}  
\small\caption{Posterior distributions of the binary star parameters for KIC\,8164262, where $i$ is the inclination of the binary star orbit in degrees; $\omega$ is the argument of periastron in radians; $ecc$ is the eccentricity; $\phi$ is the orbital phase shift; $q$ is the mass ratio ($m_2/m_1$); $sma$ is the semi-major axis in \Rsun; $vga$ is the gamma velocity in $\kms$; $logg1$ is the surface gravity in dex; and $amp1$ and $phase1$ are the amplitude (in mmag) and phase of the high-amplitude pulsation. The best fit values and uncertainties are given at the top of each column. The layout is analogous to that in Fig.\,\ref{fig:tdcr}.} 
\label{fig:posteriors} 
\hfill{} 
\end{figure*} 

The posteriors for all parameters shown in Fig.~\ref{fig:posteriors}, are well approximated by Gaussians. The list of adjustable parameters and their final values from our \mcmc\ model fit are presented in Table\,\ref{tab:free}. The parameters are indicative of a slightly evolved F star primary component. The light ratio provides an upper estimate of the secondary component's light contribution, which is suggestive of a main sequence M star. From the parameters obtained, the stars are 12.1(2)\,$\Rsun$ apart at periastron and 200(4)\,$\Rsun$ apart at apastron. This significant variation in separation is the driving force of the tidally induced pulsations observed in KIC\,8164262.

\subsection{Stellar Rotation}
\label{sec:rotation}

After pre-whitening the light curve by our orbital model, we identified two peaks in the amplitude spectrum ($\nu$\,=\,0.3345\,\cd\ and $\nu$\,=\,0.6690\,\cd) that are not orbital harmonics. The second peak is the harmonic of the first, which suggests that the peaks may be formed by rotational variations in the light curve due to spots. Furthermore, both the amplitude and phase of the peaks were found to change over the duration of the data set, consistent with spot activity. \citet{Zimmerman2017} identified 24 heartbeat stars with similar harmonic features, all of which were attributed to rotation. We thus obtain a rotational period of 2.98942(6)\,d. This is significantly faster than the pseudo-synchronous period, 4.6(2)\,d \citep{Hut1981}, in-line with the findings of \citep{Zimmerman2017}. In our binary star model we fixed the rotational period of both components to the value of 2.98942(6)\,d or $F = 29.2548(6)$, where $F$ is ratio of the stellar rotation period to the orbital period. 

Currently, we are unable to discern which star has the spots that are providing the rotational signal. As M stars are known to have significant spot coverage, even though the M star is extremely faint, it could present variations on the order of 40\,ppm. However, if we assume that the spots are on the primary star, considering the spot rotation period combined with the primary star radius determined in \S\,\ref{sec:ph} ($R_1$\,=\,2.4(1)\,$\Rsun$) and $v \sin i$ determined through spectra ($v \sin i$\,=\,23(1)\,$\kms$), we infer that the inclination of the primary star would be $i= 35(3)$\degs, compared to the orbital inclination of $incl = 65(1)$\degs. This suggests that if the spots originate on the primary star, the sky-projected inclination angle difference is, $\lambda$, is 30(3)\degs. Similar angles have been observed in objects such as DI Her, which is also a detached, eccentric ($e = 0.489$) binary system \citep{Albrecht2009} and CV Velorum, which is slightly more evolved, but shows an obliquity of $\sim$65\degs \citep{Albrecht2014}. However, we cannot rule out that the spots originate from the secondary, for which we are unable to calculate the sky-projected inclination angle difference ($\lambda$).

\section{Pulsation Characteristics}
\label{sec:pulse}

\begin{table} 
\caption{ 
\label{tab:pulse} 
\small Frequencies extracted from the masked light curve of KIC\,8164262. The majority of the frequencies extracted are multiples of the orbital frequency, with the exception of the two rotation peaks and three frequencies under 1\,d$^{-1}$. The asterisks denote that the frequency extracted is not resolved from the large amplitude 229$^{th}$ orbital harmonic. The phase is relative to the time of periastron (2455668.7898(2)). The values in parentheses denote the uncertainty in the last digits of the previous value.} 
\begin{center} 
\begin{tabular}{|l|c|r|r|} 
\hline 
Freq &Notes&Amp	&	Phase 		\\
(c\,d$^{-1}$)&&	(ppm)		&	(rad)	\\\hline
2.6184922(3)&229\,$\nu_{orb}$	&	1010(20)&	2.844(1)\\
0.334512(7)&rotation		&	41.0(8)	&	2.40(2)\\
2.755699(9)&241\,$\nu_{orb}$	&	35.3(8)	&	-1.74(2)\\
1.40645(1)&123\,$\nu_{orb}$	&	22.9(9)	&	-2.51(3)\\
2.61912(2)&229\,$\nu_{orb}$*	&	15(2)	&	-0.40(9)\\
1.80665(2)&158\,$\nu_{orb}$	&	15.2(8)	&	2.08(5)\\
1.41786(2)&124\,$\nu_{orb}$	&	15.1(9)	&	-0.93(6)\\
1.50933(2)&132\,$\nu_{orb}$	&	13.3(9)	&	-1.00(6)\\
0.66907(2)&rotation		&	12.4(8)	&	-2.70(7)\\
2.21831(2)&194\,$\nu_{orb}$	&	12.3(8)	&	1.23(7)\\
1.46360(3)&128\,$\nu_{orb}$	&	11.8(9)	&	2.63(7)\\
2.61832(3)&229\,$\nu_{orb}$*	&	11(1)	&	2.7(1)\\
3.62472(3)&317\,$\nu_{orb}$	&	9.5(8)	&	2.41(8)\\
1.47501(4)&129\,$\nu_{orb}$	&	8.3(9)	&	-0.7(1)\\
0.28033(4)&		--	&	8.3(8)	&	-2.0(1)\\
0.28383(4)&		--	&	7.6(8)	&	0.7(1)\\
1.42931(4)&125\,$\nu_{orb}$	&	6.9(9)	&	-0.5(1)\\
1.56644(4)&137\,$\nu_{orb}$	&	6.8(8)	&	2.3(1)\\
0.28504(5)&		--	&	6.6(8)	&	-0.9(1)\\
1.30355(5)&114\,$\nu_{orb}$	&	6.4(8)	&	-1.5(1)\\
2.61901(5)&229\,$\nu_{orb}$*	&	6(1)	&	2.9(2)\\
3.01870(6)&264\,$\nu_{orb}$	&	5.6(8)	&	-3.1(1)\\
0.25121(5)&22\,$\nu_{orb}$	&	5.6(8)	&	-1.7(1)\\
\hline                                  	
\end{tabular}                           	
\end{center}                            	
\end{table}

\begin{figure*}
\hfill{} 
\centerline{\includegraphics[height=10cm]{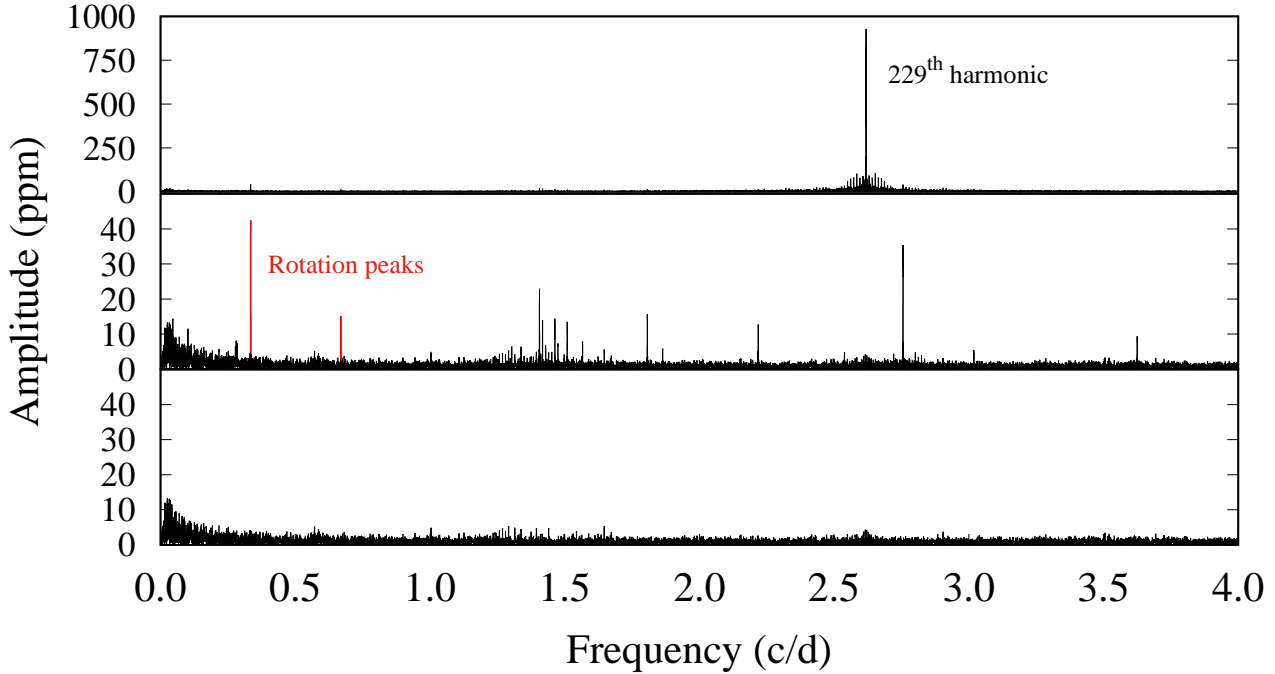}}  
\small\caption{Amplitude spectra showing the frequency spectrum at different stages. Starting from the top, depicted are the amplitude spectra with: no peaks removed (note the single prominent peak at 2.6\cd, the 229$^{th}$ orbital harmonic); the binary model and primary pulsation subtracted (note the change of scale), where the two rotation peaks at 0.3345\,\cd\ and 0.6690\cd\ are highlighted in red; all peaks removed to an amplitude of 4\,$\mu$mag.} 
\label{fig:ft} 
\hfill{} 
\end{figure*} 

The light curve of KIC\,8164262 contains one high-amplitude, tidally-excited mode ($\nu$ = 229\,$\nu_{orb}$, cf. Fig.\,\ref{fig:lc_model}), which we fitted simultaneously with the binary star model. The amplitude and phase of the 229$^{th}$ orbital harmonic were found to be $A = 1.01(2)$\,mmag and $\phi = 5.19(6)$\,rad relative to the zero point specified in \S\,\ref{sec:period} or $\phi = 2.844(1)$ relative to periastron. In Fig.\,\ref{fig:ft}, amplitude spectra with no peaks removed (top panel), with the binary model and 229$^{th}$ orbital harmonic subtracted (middle panel) and with all the peaks to an amplitude of 4\,$\mu$mag removed can be seen. The stunning prominence of the 229$^{th}$ orbital harmonic can be seen in the top panel (note the change of scale for the different panels).

The binary star model, including the high amplitude mode, was then subtracted from the time-series light-curve data and the residuals were analysed. To remove any residual information from the binary star features, the data points were removed from the region between phases -0.01 to 0.01, the phases of the ellipsoidal variation in the original time series. In the amplitude spectra, these gaps appear in the window pattern, separated from the main peak by the orbital frequency. However, as the binary orbital period is long compared to the duration of the ellipsoidal variation, the removal of these points did not create a window pattern with significant additional amplitude in the sidelobes. We calculated amplitude spectra to these data and found that the highest amplitude pulsation peak that remained (after the significant high amplitude peak had been removed) had an amplitude of $35$\,ppm, which is 3.5 per cent of the dominant pulsation.

We individually extracted each peak from the amplitude spectra until we reached an amplitude of 5\,$\mu$mag. Beyond this point we were unable to distinguish between pulsation frequencies and noise. Table\,\ref{tab:pulse} provides a list of the extracted frequencies, amplitudes and phases relative to periastron. We identified three peaks that are unresolved from the prominent 229th orbital harmonic. It is likely that these are due to phase and/or amplitude variation, which is common in \DS\ stars \citep{Bowman2014, Bowman2016}. We also provide the orbital harmonic number for each pulsation frequency -- all peaks are orbital harmonics with the exception of three at $\nu = 0.28033(4)$\,\cd, $\nu = 0.28383(4)$\,\cd and $\nu = 0.28504(5)$\,\cd, and the two rotational peaks at $\nu =$\,0.3345\,\cd\ and $\nu =$\,0.6690\,\cd, discussed in \S\ref{sec:rotation}. We speculate that the former three pulsations are either the unresolved rotation signal of the secondary component; or naturally occurring g mode pulsations, as the primary star temperature is consistent with that of a \GD\ pulsator, which pulsates on the order of 1\,\cd \citep{Grigahcene2010a}.

We extended our frequency search beyond the Nyquist frequency to ensure that we had identified all the peaks and that the selected peaks were not Nyquist aliases. All peaks beyond the Nyquist frequency showed a multiplet structure caused by the irregular time sampling of \kep\ data due to satellite motion \citep{Murphy2013}. Thus we conclude that the identified peaks are the real peaks.

\section{Comparison with Stellar Evolutionary Models}

To assess the consistency of our models and to ascertain the age of the primary, pulsating star, we generated evolutionary models and isochrones and compared them with the results for the primary component. This was not possible for the secondary component, as we were unable to determine its radius and effective temperature due to its low luminosity (relative to the primary component). We used the {\sc {mist}} ({\sc {mesa}} Isochrones and Stellar Tracks) software \citep{Dotter2016,Choi2016}, which is based on the {\sc {mesa}} (Modules for Experiments in Stellar Astrophysics) package \citep{Paxton2011,Paxton2013,Paxton2015}. We generated evolutionary models with [Fe/H]\,=\,0.5, the metalicity determined when fitting the KPNO spectra (see \S\ref{sec:todcor}). The models incorporated time-dependent, diffusive convective overshoot \citep{Herwig2000}, \citet{asplund2009} solar abundances and the OPAL \citep{Iglesias1993, Iglesias1996} opacities.

\begin{figure}
\centerline{\includegraphics[height=6cm]{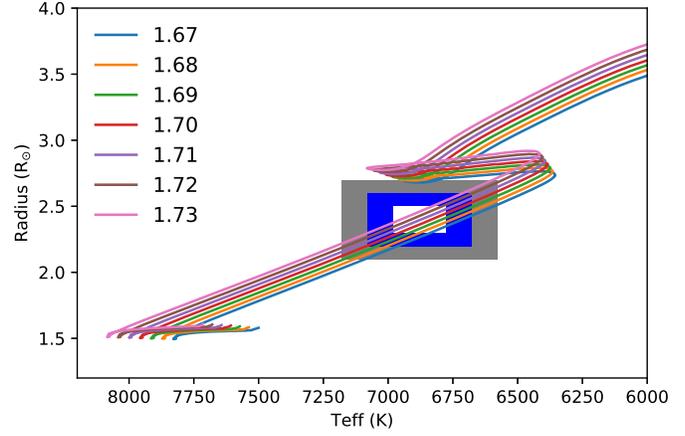}}  
\small\caption{Main-sequence and post-main-sequence evolutionary tracks from the {\sc {mist}} series for the measured mass of the primary component of KIC\,8164262 in the Teff--radius plane. The lines represent evolutionary models in one sigma increments about the central mass value of $M_1 = 1.7 \pm 0.1 \Msun$. The boxes represent the 1$\sigma$ (white), 2$\sigma$ (blue) and 3$\sigma$ (grey) results from the binary model fit. The legend provides the mass for each evolutionary track in units of solar mass.}
\label{fig:evo} 
\end{figure}

Fig.\,\ref{fig:evo} shows the evolutionary tracks for stars ranging from 1.63 to 1.73\,$\Msun$ in steps of 0.01\,$\Msun$, which bracket the results of the binary model fit ($M_1 = 1.7 \pm 0.1$) in steps of $1\sigma$. The boxes represent the one, two and three\,$\sigma$ results from the model fit in the Teff--radius plane. 


\begin{figure}
\centerline{\includegraphics[height=6cm]{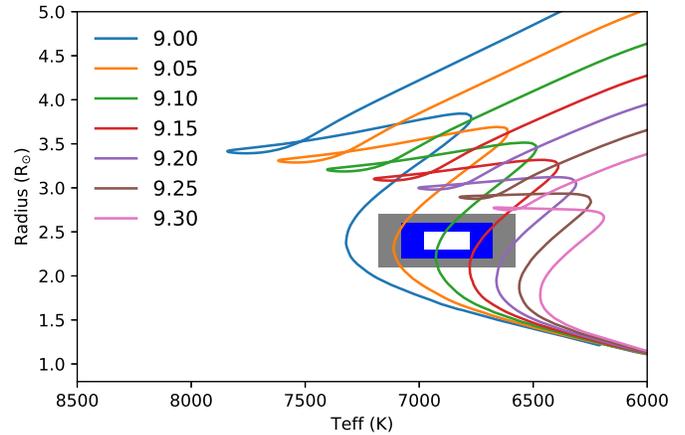}}  
\small\caption{Stellar isochrones in the Teff--radius plane to determine the age of the primary component of KIC\,8164262. The lines represent log(Age[yr]) and the legend provides the ages in log(Age[yr]). The best matching age is log(Age[yr])\,=\,9.1 or 1.2\,Gyr. The boxes represent the 1$\sigma$ (white), 2$\sigma$ (blue) and 3$\sigma$ (grey) results from the binary model fit. }
\label{fig:iso} 
\end{figure}

To determine the age of the primary component, we compared the binary model results with stellar isochrones, again with [Fe/H]\,=\,0.5. As seen in Fig.\,\ref{fig:iso}, the models suggest the primary component is log(Age[yr])\,=\,9.1 or Age\,=\,1.2\,Gyr. The models also show that the primary component is nearing the end of the main sequence.

\section{Discussion and Conclusions}
\label{sec:summary}

KIC\,8164262 is an extreme \hb\ star, in that its orbital period and eccentricity are larger than most of the heartbeat stars discovered so far. Its most striking feature is the high amplitude ($\sim$1\,ppt) tidally excited pulsation at 229 $\nu_{\rm orb}$ --- the largest amplitude tidally excited pulsation observed to date. The frequency of this pulsation is not unusual (frequencies of $0.5 \, {\rm d}^{-1} \lesssim \nu_{\rm orb} \lesssim 3 \, {\rm d}^{-1}$ are common in \hb\ stars), it simply occurs at a large orbital harmonic because of the small orbital frequency. However, the amplitude of the pulsation is exceptional, as it is over twenty times larger than any other pulsation in KIC\,8164262, and roughly four times larger than any pulsations in KOI-54.  

The LC \kep\ data of Quarters 0--17 and radial velocities from three different telescopes (Keck, the 4-m Mayall telescope at KPNO and the 2.7-m telescope on the McDonald Observatory) were modelled using \ph\ combined with \emcee. We further augmented the software to add sine waves to the light curve to model the tidally induced pulsations, which we used to model the prominent tidally induced pulsation at 229\,$\nu_{orb}$, and to model Doppler boosting. The results of the spectral analysis on the KPNO spectra, specifically the effective temperature and log\,$g$ of the primary component, were also incorporated into the modelling effort to fully constrain the fundamental parameters.  Using these combined software packages, we determined that KIC\,8164262 contains a slightly evolved F star, which is experiencing the tidally induced pulsations, and a tentatively classified M dwarf star. Comparing the model results with MIST stellar evolution tracks and isochrones confirms the primary is approaching the end of the main sequence and suggests an age of 1.2 Gyr.

We performed pulsational analysis on the complete \kep\ light curve with the binary star model removed. We found that all the identified peaks, with the exception of five, were multiples of the orbital frequency, thus we conclude that these are all tidally induced pulsations. Of the remaining 5 we identified three peaks at $\nu = 0.28033(4)$\,\cd, $\nu = 0.28383(4)$\,\cd and $\nu = 0.28504(5)$\,\cd, which are likely g-mode pulsations. The remaining two peaks have frequencies at $\nu =$\,0.3345\,\cd\ and $\nu =$\,0.6690\,\cd, where the latter is the harmonic of the former, suggestive of rotational variations due to spots \citep{Zimmerman2017}.

The a large amplitude mode requires explanation and may yield clues to tidal dissipation processes in binary star systems. In a companion paper, F17, Fuller et al. calculate the expected frequencies and amplitudes of tidally excited pulsations from theoretical considerations. Fuller et al. find that an extremely finely tuned resonance is required to tidally excite a mode to the observed amplitude, and such a resonance is unlikely to occur by chance. Instead, they find that the pulsation is well explained (in both frequency and amplitude) as a resonantly locked mode. In this scenario, the combined effects of stellar evolution and spin-down are balanced by ongoing tidal circularization and synchronization in a self-regulating process such that the frequency of a single pulsation mode is held at resonance with the tidal forcing. The result is an increased rate of tidal dissipation compared to conventional expectations (see \citet{Zahn2008} for a review). For A--F stars, tidal interactions are expected to be weak due to the absence of a thick convective envelope and the presence of only a small convective core, entailing an effective tidal quality factor (which measures the efficiency of tidal dissipation) of $ Q \gtrsim 10^6$. However, for KIC\,8164262 F17 calculate the effective tidal quality factor to be $Q \sim 5\times 10^4$ while the resonance lock is active, corresponding to an orbital circularization timescale of $\sim 5\, {\rm Gyr}$. Furthermore, in the absence of the prominent pulsation, F17 finds that the circularization timescale is $\sim500$ times longer. This is suggestive of the importance of resonance locking for the acceleration of orbital circularization. Further details are presented in F17.

\section{Acknowledgments}
The authors express their sincere thanks to NASA and the \kep\ team for the high quality \kep\ data. The \kep\ mission is funded by NASA's Science Mission Directorate. We thank the Planet Hunters for finding this object and Professor Dan Fabrycky for notifying us about it. This work was also supported by the STFC (Science and Technology Funding Council). KH, ST and JF acknowledge support through NASA K2 GO grant (11-KEPLER11-0056) and NASA ADAP grant (16-ADAP16-0201). AP and KH acknowledge support from the NSF (grant \#1517460).  JF acknowledges partial support from NSF under grant nos. AST-1205732,  PHY-1125915, and through a Lee DuBridge Fellowship at Caltech. We would like to thank the RAS for providing grants which enabled KH's attendance to conferences and thus enabled the development of collaborations and the successful completion of this work. AP acknowledges support through NASA K2 GO grant (NASA 14-K2GO1\_2-0057). We acknowledge the observations taken using the 4-m Mayall telescope, KPNO (NOAO 2011A-0022); the Keck telescope, Mauna Kea; and the 2.7-m telescope at the McDonald Observatory. Some of the data presented in this paper were obtained from the Mikulski Archive for Space Telescopes (MAST). STScI is operated by the Association of Universities for Research in Astronomy, Inc., under NASA contract NAS5-26555. Support for MAST for non-HST data is provided by the NASA Office of Space Science via grant NNX09AF08G and by other grants and contracts.

\bibliographystyle{mn2e}
\bibliography{StarRef_2}  

\end{document}